\documentstyle[preprint,tighten,pra,aps,epsf]{revtex}
\begin{document}
\tightenlines
\draft

\preprint{UTF 412}

\title{Quantum Properties of Topological Black Holes}

\author{Dietmar Klemm\footnote{email: klemm@science.unitn.it}}
\address{Dipartimento di Fisica, Universit\`a  di Trento, Italia}
\author{Luciano Vanzo\footnote{email: vanzo@science.unitn.it}}
\address{Dipartimento di Fisica, Universit\`a di Trento\\ 
and Istituto Nazionale di Fisica Nucleare\\ 
Gruppo Collegato di Trento, Italia}

\maketitle\maketitle\begin{abstract}
We examine quantum properties of topological black holes which
are asymptotically anti--de Sitter. First, massless scalar fields
and Weyl spinors which propagate in the background of an anti--de Sitter
black hole are considered in an exactly soluble two--dimensional toy model.
The Boulware--, Unruh--, and Hartle--Hawking vacua are defined. The latter
results to coincide with the Unruh vacuum due to the boundary conditions
necessary in asymptotically adS spacetimes. We show that the Hartle--Hawking
vacuum represents a thermal equilibrium state with the temperature found
in the Euclidean formulation. The renormalized stress tensor for this
quantum state is well--defined everywhere, for any genus and for all solutions 
which do not have an inner Cauchy horizon, whereas in this last case it 
diverges on the inner horizon.
The four--dimensional case is finally considered, 
the equilibrium states are discussed and a luminosity formula for the
black hole of 
any genus is obtained. Since spacelike infinity in anti--de Sitter 
space acts like a mirror, it is pointed out how this would imply 
information loss in gravitational collapse. The black hole's mass spectrum 
according to Bekenstein's view is discussed and compared to that provided by 
string theory. 
\end{abstract}


\maketitle

\section{Introduction}

Since the discovery of black holes whose event horizons have
nontrivial topology \cite{amin97,mann97,vanz97,zhang} there has been much
research activity in this area. Charged versions of these
black holes were presented in \cite{mann97,zhang}, and also rotating
generalizations are known \cite{lemzan,kmv97}.
Moreover, they also exist in dilaton gravity \cite{cai}.
Mann \cite{manW97} and
Lemos \cite{lemo97} showed that topological black holes
can form by gravitational collapse. Finally, from a thermodynamical point of
view, they are well--behaved objects
obeying the entropy--area law \cite{vanz97,bril97}.\\ 
In this paper we want to investigate questions such as particle 
production by the black holes, and their equilibrium state, the 
Israel--Hartle--Hawking quantum state familiar in the Schwarzschild 
case \cite{isra76,hart76,hawking,wald}. As the local temperature
of the black holes 
is zero at infinity due to a diverging lapse
function (infinite redshift), it is possible that there is no net flux 
of radiated particles at infinity, and that therefore
no Unruh--like states exist.\\ 
We will show that this is indeed the case, and that one can 
define equilibrium thermal states. Part of the interest in doing so is one 
important feature that distinguishes anti--de Sitter from asymptotically flat 
black holes, namely the asymptotic behaviour at infinity. The 
timelike character of the boundary of the space manifests itself in the 
black hole spacetime by making the exterior static region non--globally 
hyperbolic. This is remedied by imposing boundary conditions 
at spacelike infinity \cite{avis78,brei82}, to 
prevent radiation from escaping out. As a result, 
the particles emitted by the black hole will be ultimately recaptured  
back and the black hole evolution will not be complete evaporation. 
Hence one suspects that the final state should be a thermal 
equilibrium state, implying a maximum of information loss. \\ 
In fact, one can imagine a pure state collapsing in anti--de Sitter 
space to end into a mixture with the largest entropy available. Of 
course, this is due to infinity in anti--de Sitter acting like a 
boundary, which seems a rather artificial situation. However, it is an 
important matter of principle showing that gravity can imply 
information loss, since now there is no point for the information to 
return, apart from the existence of white holes \cite{hawk76}.\\ 
Another point of interest is the discrete quantization of the
one--particle energies in anti--de Sitter space. When a black hole is 
present, this is no longer true and a continuous spectrum appears. The 
density of states is then blue shifted to infinity near the horizon, so one 
cannot explain the black hole entropy as resulting from entangled 
states, without invoking a short distance cut--off \cite{thoo85} (see 
\cite{beke94,frol98} for recent reviews). \\ 
Finally, there is the question of the black hole mass spectrum in the 
spirit of Bekenstein \cite{mukh86,beke}. To match with the 
Euclidean partition function, we obtain a mass spectrum $M_n=\sigma 
n^{3/2}$, which seems to be difficult to conciliate with string theory. 
However, recently the entropy of three-- and five--dimensional anti--de Sitter 
black holes has been explained by string theory \cite{birm98}. Even 
more interesting is the agreement between entropy and the degeneracy of 
states in a conformal field theory \cite{stro97,bana98} realizing 
the asymptotic symmetries of three--dimensional anti--de Sitter gravity 
\cite{brow86}. These important new developments may have an impact on 
the more difficult four--dimensional case.\\ 
The rest of the paper is structured as follows:\\ 
We begin in section (II) by presenting the geometry of the
problem.\\ 
In section (III) we consider massless scalar fields and Weyl
spinors propagating in a two--dimensional black hole background.
The Boulware--, Unruh--, and Hartle--Hawking vacua are defined.
We then calculate the renormalized stress tensor for the three
states and show that in the Hartle--Hawking case it indeed describes
a thermal equilibrium state. Furthermore we will see that,
unlike the Sch\-warz\-schild black hole, the Unruh--
and the Hartle--Hawking states coincide, and that this can be traced
back to the boundary conditions which have to be imposed at infinity,
as our spacetime fails to be globally hyperbolic. Though the
absence of a true Unruh state in its original meaning, we show
that a black hole luminosity can be defined in a certain sense,
and calculate it
using the Bogoljubov coefficients relating the Unruh-- and the Boulware
vacuum.\\ 
In section (IV) we consider the more complicated
four--di\-mens\-ion\-al case, in which backscattering is present. We 
argue that the most likely final state is a thermal 
Israel--Hartle--Hawking state. This implies information loss.\\ 
In section (V) we address the question of the mass spectrum from the 
simple perspective of Bekenstein--Mukhanov arguments. We discuss this spectrum 
from the Susskind--Horowitz--Polchinski point of view and we point out 
the difficulty to explain the black hole entropy with it, at least naively.
At last, our results will be summarized and discussed in section (VI).\\ 
In this paper we shall use the curvature conventions of Hawking--Ellis'
book \cite{haell} and employ Planck's dimensionless units.

\section{Spacetime Geometry} \label{spacegeo}
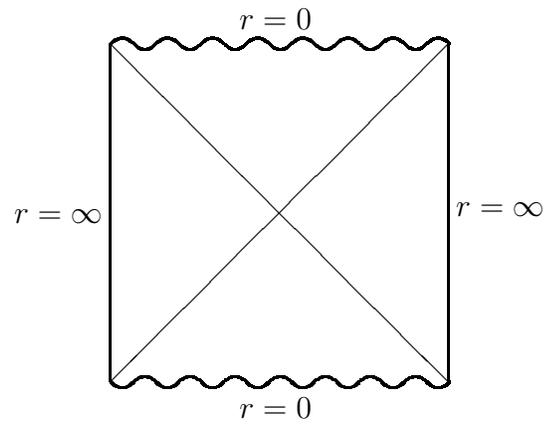
\begin{figure}
\vglue 1 cm

\unitlength 1.00mm
\linethickness{0.4pt}
\begin{picture}(106.00,57.00)
\put(105.00,55.00){\line(-1,-1){45.00}}
\put(60.00,55.00){\line(1,-1){45.00}}
\thicklines
\put(105.00,10.00){\line(0,1){45.00}}
\put(60.00,10.00){\line(0,1){45.00}}
\bezier{20}(105.00,55.00)(103.50,56.50)(102.00,55.00)
\bezier{20}(102.00,55.00)(100.50,53.50)(99.00,55.00)
\bezier{20}(99.00,55.00)(97.50,56.50)(96.00,55.00)
\bezier{20}(93.00,55.00)(91.50,56.50)(90.00,55.00)
\bezier{20}(87.00,55.00)(85.50,56.50)(84.00,55.00)
\bezier{20}(81.00,55.00)(79.50,56.50)(78.00,55.00)
\bezier{20}(75.00,55.00)(73.50,56.50)(72.00,55.00)
\bezier{20}(69.00,55.00)(67.50,56.50)(66.00,55.00)
\bezier{20}(63.00,55.00)(61.50,56.50)(60.00,55.00)
\bezier{20}(96.00,55.00)(94.50,53.50)(93.00,55.00)
\bezier{20}(90.00,55.00)(88.50,53.50)(87.00,55.00)
\bezier{20}(84.00,55.00)(82.50,53.50)(81.00,55.00)
\bezier{20}(78.00,55.00)(76.50,53.50)(75.00,55.00)
\bezier{20}(72.00,55.00)(70.50,53.50)(69.00,55.00)
\bezier{20}(66.00,55.00)(64.50,53.50)(63.00,55.00)
\bezier{20}(105.00,10.00)(103.50,8.50)(102.00,10.00)
\bezier{20}(102.00,10.00)(100.50,11.50)(99.00,10.00)
\bezier{20}(99.00,10.00)(97.50,8.50)(96.00,10.00)
\bezier{20}(93.00,10.00)(91.50,8.50)(90.00,10.00)
\bezier{20}(87.00,10.00)(85.50,8.50)(84.00,10.00)
\bezier{20}(81.00,10.00)(79.50,8.50)(78.00,10.00)
\bezier{20}(75.00,10.00)(73.50,8.50)(72.00,10.00)
\bezier{20}(69.00,10.00)(67.50,8.50)(66.00,10.00)
\bezier{20}(63.00,10.00)(61.50,8.50)(60.00,10.00)
\bezier{20}(96.00,10.00)(94.50,11.50)(93.00,10.00)
\bezier{20}(90.00,10.00)(88.50,11.50)(87.00,10.00)
\bezier{20}(84.00,10.00)(82.50,11.50)(81.00,10.00)
\bezier{20}(78.00,10.00)(76.50,11.50)(75.00,10.00)
\bezier{20}(72.00,10.00)(70.50,11.50)(69.00,10.00)
\bezier{20}(66.00,10.00)(64.50,11.50)(63.00,10.00)
\thinlines
\put(106.00,33.00){\makebox(0,0)[lc]{$r=\infty$}}
\put(59.00,32.00){\makebox(0,0)[rc]{$r=\infty$}}
\put(82.00,57.00){\makebox(0,0)[cb]{$r=0$}}
\put(82.00,8.00){\makebox(0,0)[ct]{$r=0$}}
\end{picture}

\vglue 1 cm
\caption{Penrose--Carter diagram for the toroidal black hole with
$\eta > 0$.}
\label{torus}
\end{figure}

All black hole metrics we shall consider are of the form
\begin{equation}
ds^2=-V(r)dt^2+V^{-1}(r)dr^2+r^2d\sigma^2,
\label{metri}
\end{equation}
where $d\sigma^2$ is the metric of a constant curvature Riemann surface 
with genus $g$, denoted $S_g$. For $g=1$ we shall take this metric to 
be
\begin{equation}
d\sigma^2=dx^2+|\tau|^2dy^2+2\mbox{Re}\tau dx\,dy\hspace{1cm} 
\mathop{\rm Im}\nolimits\tau>0,
\label{tmet}
\end{equation}
where $(x,y)$ ranges over the closed unit square $[0,1]\times[0,1]$. 
The element $\tau$ is known as the Teichm\"uller parameter of the 
torus. The lapse $V(r)$ is given by
\begin{equation}
V(r) = -1 + \delta_{g,1} -\frac{2\eta}{r} + \frac{r^2}{\ell^2},
\end{equation}
where $\Lambda = -3\ell^{-2}$ is the cosmological constant. Denoting with 
$r_+$ the position of the horizon (the outermost zero of the lapse), the 
surface gravity of the black holes is
\begin{eqnarray}
\kappa=\frac{3r_+^2-\ell^2}{2r_+\ell^2}, \qquad \kappa=\frac{3r_+}{2\ell^2}
\label{surfgrav}
\end{eqnarray}
for $g>1$ and $g=1$ respectively. As will be seen, the quantity 
$T=\kappa/2\pi$ is the Hawking temperature of the black hole.\\
For $g>1$, the parameter $\eta$ is related to the mass
of the black hole by $M=(\eta+\ell/3\sqrt{3})(g-1)\geq0$, and for
$g=1$ by $M=\eta|\mathop{\rm Im}\nolimits\tau|/4\pi\geq0$ \cite{vanz97}. The 
positivity condition for the mass means that the solutions have a black 
hole interpretation for $\eta>-\ell/3\sqrt{3}(1-\delta_{g,1})$, the other 
values giving naked singularities or extreme black holes, in general. \\
The reason for this connection between the ADM mass and the parameter 
$\eta$, in the higher genus case, can be understood by finding
the zero temperature state. 
For a toroidal black hole, the temperature vanishes precisely when 
$\eta=0$, but in the $g>1$ case, the zero temperature state has 
$\eta=-\ell/3\sqrt{3}$. This represents an extremal black hole, where the 
Cauchy inner horizon has merged with the event horizon, and is to be 
considered as the ground state.\\ 
The Euclidean section of the extreme state can be identified to any 
period, without introducing conical singularities at the origin. We may 
then compute the Euclidean action of a black hole relative to the 
extreme state, and define the mass as the thermal energy in the 
canonical ensemble corresponding to the Hawking temperature. As has 
been shown in \cite{vanz97}, this yields
\begin{eqnarray}
M\!&=&\!\frac{(g-1)4\pi^3\ell^4T^3}{27}\left(1+\sqrt{1+
\frac{3}{4\pi^2\ell^2T^2}}\right)\left(2-\frac{3}{2\pi^2\ell^2T^2}+
2\sqrt{1+\frac{3}{4\pi^2\ell^2T^2}}\right)\nonumber \\
&+&\left(\frac{\ell}{3\sqrt{3}}\right)(g-1).
\label{mass}
\end{eqnarray}
This mass is an increasing function of $T$ in the full range 
$0\leq T\leq\infty$, with a large-$T$ behaviour $M\sim T^3$. Any other 
choice for the background gives a negative contribution to the entropy 
of the black hole, because one cannot identify the solution and the 
background with the same temperature without introducing conical
singularities, except for the extreme state. In other words, one cannot 
shift the mass and leave unaffected the entropy at the same time.\\
By expressing the mass in (\ref{mass}) in terms of $\eta$, one finds 
the relation we started with, $M=(\eta+\ell/3\sqrt{3})(g-1)$. One can 
also show that this is precisely the on-shell value of the Hamiltonian, 
relative to the zero temperature state.\\
Let us restrict our considerations to
the case $g=1$, i.~e.~to the torus. The generalization to $g > 1$ can be
done in a straightforward way, if the parameter $\eta>0$ is positive.
The fact that $\eta$ can become
negative in the higher genus case, and an inner horizon forms for
$\eta<0$, has many implications for the results we will
obtain, but will not affect the discussion of the Hawking radiation 
perceived by an external observer. Instead, global classical and quantum 
properties, such as global hyperbolicity and the
existence of the Hartle-Hawking 
state, will be affected heavily. 
Throughout this paper, we will indicate the modifications
of our results in the $g>1$ case at the appropriate places.
For our subsequent discussion it will be convenient to introduce several
other coordinate systems. First of all, define the so--called tortoise
coordinate $r_*$ by
\begin{eqnarray}
r_* &=& \int\frac{dr}{V(r)} \nonumber \\ 
    &=& \frac{\ell^2}{r_+}\left[\frac{1}{6}\ln\frac{(r-r_+)^2}{r^2 + rr_+
        + r_+^2} + \frac{1}{\sqrt{3}}\arctan\frac{2r + r_+}{r_+\sqrt{3}}
        - \frac{\pi}{2\sqrt{3}}\right],
\end{eqnarray}
where $r_+ = (2\eta\ell^2)^{1/3}$ is the location of the event horizon and
the integration constant is chosen so that $r_* \to 0$ for $r \to \infty$.
Then introduce retarded/advanced null coordinates $u,v$ according to
\begin{equation}
u = t - r_*, \hspace{1cm} v = t + r_*.
\end{equation}
In these coordinates the line element reads
\begin{equation}
ds^2 = - V(r)dudv + r^2 d\sigma^2.
\end{equation}
Finally, Kruskal coordinates $U,V$ are defined by
\begin{equation}
U = -\exp(-\kappa u), \hspace{1cm} V = \exp(\kappa v),
\end{equation}
with $\kappa$ given by (\ref{surfgrav}).
In Kruskal coordinates we obtain for the metric
\begin{equation}
ds^2 =  \frac{V(r)}{\kappa^2 UV}\,dUdV + r^2 d\sigma^2,
\end{equation}
which is regular on the horizon. Figure \ref{torus} shows the conformal
diagram of the spacetime.\\ 
It is an important fact that the solutions so far discussed may appear 
as a result of gravitational collapse of a configuration of pressureless dust 
\cite{manW97}. The exterior metric matches with an interior 
Robertson--Walker spacetime, and complete collapse occurs in a finite 
amount of co--moving time. The fact that the exterior metric is given by 
(\ref{metri}) at all times results from a generalization of Birkhoff's
theorem \cite{bron80}.  

\section{The Two--Dimensional Case}

In our two--dimensional toy model 
we forget the part $r^2d\sigma^2$ in (\ref{metri}), i.~e.~we limit
the calculations to the toroidal analogue of the spherical s--wave sector.
This restriction "does not throw out the baby with the water",
as expressed A.~Strominger \cite{strom}. Indeed, we shall see that most of
the essential features are present in this model. In two dimensions,
there is no backscattering, the metric is conformally flat, and the
Klein--Gordon or the Weyl equation can be solved exactly.\\ 

\subsection{Propagation of Massless Scalar Particles}
Since the metric is conformally flat, the
Klein--Gordon equation for a massless field $\phi$ is just the same as in
Minkowski space, namely
\begin{equation}
\frac{\partial^2 \phi}{\partial u \partial v} = 0, \quad \mbox{or}
\quad \frac{\partial^2 \phi}{\partial U \partial V} = 0,
\end{equation}
which has the general solution
\begin{eqnarray}
\phi(u,v) &=& f(u) + g(v) \nonumber \\ 
\mbox{or} \quad \phi(U,V) &=& F(U) + G(V).
\end{eqnarray}
$f,g,F,G$ are arbitrary functions. Now in the
Unruh model \cite{Unruh} of an evaporating Schwarzschild black hole,
one defines e.~g.~a complete set of Boulware modes by
$\varphi_{in}^{B} \propto
\exp(-i\omega v)$
and $\varphi_{out}^{B}
\propto \exp(-i\omega u)$, which means
that we have neither particles incoming from past infinity
nor particles outgoing to future infinity in the corresponding vacuum state.
Analogously, Hartle--Hawking (HH) modes
are defined by $\varphi_{in}^{HH}
\propto \exp(-i\omega U)$ and
$\varphi_{out}^{HH} \propto
\exp(-i\omega V)$,
i.~e.~the Hartle--Hawking vacuum
does not contain Kruskal particles emerging from the white hole horizon $H^-$
or crossing the future horizon $H^+$. Finally, Unruh modes are given by
$\varphi_{in}^{U} \propto
\exp(-i\omega U)$ and
$\hat{\varphi}_{in}^{U} \propto
\exp(-i\omega v)$. Now this procedure
is no more possible in the case of topological black holes.
The Penrose diagram in figure \ref{torus} shows that the spacetime
is not globally hyperbolic; in order to obtain a well--posed Cauchy
problem we have to impose boundary conditions at infinity.
There are three natural choices of boundary conditions at $r = \infty$
(which is equivalent to $u = v$ or $UV = -1$):
\begin{eqnarray}
\phi|_{r = \infty} &=& 0 \qquad \mbox{(Dirichlet)} \nonumber \\ 
\nabla_n \phi|_{r = \infty} &=& 0 \qquad \mbox{(Neumann)} \nonumber \\ 
{[}K(r,t)\phi + \nabla_n \phi]_{r = \infty} &=& 0 \qquad \mbox{(Robin)}
\label{bound}
\end{eqnarray}
Here $n$ denotes the unit normal to the surface $r = \infty$, and $K$
is a function which will be given below.
Obviously the above--defined functions for the Schwarzschild case do
not fulfil any of the boundary conditions (\ref{bound}).
Therefore we define the Boulware modes to be
\begin{equation}
\varphi_{\omega}^{B}(u,v) = \frac{1}{\sqrt{4\pi\omega}}
                   (\exp(-i\omega u) \pm \exp(-i\omega v)), 
\label{Bmodes}
\end{equation}
where the minus (plus) sign corresponds to Dirichlet (Neumann) boundary
conditions. For Robin boundary conditions we take
\begin{equation}
\varphi_{\omega\,R}^{B} = \frac{1}{1 + \omega^2k^{-2}}
\left(\varphi_{\omega\,D}^B + i\omega k^{-1}\varphi_{\omega\,N}^B\right)
\label{BmodesR}
\end{equation}
with $k \in {\hbox{{\rm I}\kern-.2em\hbox{\rm R}}}$.
Here the function $K$ is simply a constant which equals
$-k$. The limit cases $k=0$ $(k \to \infty)$ represent Neumann (Dirichlet) 
boundary conditions.
The modes (\ref{Bmodes}), (\ref{BmodesR}) form a complete orthonormal set
with respect to the Klein--Gordon scalar product
\begin{equation}
(\alpha,\beta)_{KG} = i\int_{\Sigma}(\bar{\alpha}\nabla_a\beta
                             - \beta\nabla_a\bar{\alpha})n^a dx,
\end{equation}
where $\Sigma$ is a spacelike hypersurface with unit normal $n^a$, and $dx$
is the induced volume element on $\Sigma$. Thus we have
\begin{equation}
(\varphi_{\omega}^B, \varphi_{\omega'}^B) =
\delta(\omega - \omega').
\end{equation}
We can then expand the field in terms of the Boulware modes
\begin{equation}
\phi(u,v) = \int_0^{\infty}d\omega (b_{\omega}\varphi_{\omega}^B
            + b_{\omega}^{\dagger}\bar{\varphi_{\omega}}^{B}). \label{expanB}
\end{equation}
The Boulware vacuum $|B\rangle$ is now determined by
\begin{equation}
b_{\omega}|B\rangle = 0.
\end{equation}
We define Unruh modes according to
\begin{equation}
\varphi_{\omega}^{U}(U,V) = \frac{1}{\sqrt{4\pi\omega}}
(\exp(-i\omega U) \pm \exp(i\omega/V)), \label{Umodes}
\end{equation}
the +/- again denoting Neumann/Dirichlet boundary conditions respectively.
For Robin boundary conditions we have
\begin{equation}
\varphi_{\omega\,R}^{U} = \frac{1}{1 + \omega^2k^{-2}}
\left(\varphi_{\omega\,D}^U + i\omega k^{-1}\varphi_{\omega\,N}^U\right)
\label{UmodesR}
\end{equation}
with $k \in {\hbox{{\rm I}\kern-.2em\hbox{\rm R}}}$ and $K(U) = kU$.
The modes (\ref{Umodes}), (\ref{UmodesR}) resemble the solutions
of the moving mirror problem
(see e.~g.~\cite{davies,ford}). Indeed, the infinity $UV = -1$ can be
viewed as a moving mirror $V = V(U) = -1/U$, and a positive frequency
wave $e^{-i\omega U}$ outgoing from the past horizon is reflected
by the mirror and becomes a wave $e^{i\omega/V}$ travelling towards $H^+$.\\ 
On the past horizon $V=0$ only the first summand of (\ref{Umodes})
survives, since the
second oscillates infinitely fast, and hence does not give any contribution
if one constructs wave packets. The functions (\ref{Umodes}), (\ref{UmodesR})
represent again a
complete orthogonal system, i.~e.~they satisfy
\begin{equation}
(\varphi_{\omega}^{U}, \varphi_{\omega'}^{U}) =
\delta(\omega - \omega').
\end{equation}
One can also expand the field in terms of Unruh modes
\begin{equation}
\phi(U,V) = \int_0^{\infty}d\omega (a_{\omega}\varphi_{\omega}^{U}
            + a_{\omega}^{\dagger}\bar{\varphi_{\omega}}^{U}).
\end{equation}
Finally we define the Unruh vacuum $|U\rangle$ by
\begin{equation}
a_{\omega}|U\rangle = 0.
\end{equation}
Obviously $|U\rangle$ does not contain particles
emerging from the past horizon.
Last, Hartle--Hawking modes are given by
\begin{equation}
\varphi_{\omega}^{HH}(U,V) = \frac{1}{\sqrt{4\pi\omega}}
(\exp(-i\omega V) \pm \exp(i\omega/U)). \label{HHmodes}
\end{equation}
For Robin boundary conditions we take an expression corresponding to
(\ref{UmodesR}). If one constructs wave packets, only the first term
of (\ref{HHmodes}) gives a contribution on the future horizon $U=0$.
The Hartle--Hawking vacuum $|H\rangle$ is defined by the
usual procedure, analogous to the two preceding cases.
$|H\rangle$ does not contain particles
crossing the future horizon, but there are
also no particles emerging from the white hole, as we shall see later
by investigating the renormalized stress tensor.\\ 
Note that in contrast to two--dimensional adS space, where the spectrum is
discrete \cite{brei82,saka85}, in our case we have a continuous frequency
spectrum. This will also hold true in four dimensions (see section \ref{4d}).
It should not be surprising, because our spacetime is only asymptotically
locally adS, and one boundary condition at $UV = -1$ clearly does not
cause a discrete spectrum.

\subsection{Weyl Fermions}
Let us now devote attention to Weyl fermions propagating in the black hole
background. This case presents no particular difficulties, but 
there are also significant differences. The Weyl equation in a
two--dimensional
curved spacetime enjoys the conformal invariance on the same footing 
of the massless Klein-Gordon equation. We introduce a zweibein field 
$e_i^a(x)$, deserving the latin indices $i, j, k,..$ to denote local 
Lorentz tensors and $a, b, c,..$ to denote coordinate tensor 
components, and the two--dimensional Dirac matrices $\rho^i$. 
Then given a pair $(g_{ab},\psi)$ consisting of a metric tensor and a 
Weyl spinor satisfying the Weyl equation 
\begin{equation}
\rho^ie_i^a\nabla_a\psi=0,
\end{equation}
the transformed pair $\tilde{\psi}=\lambda^{-1/2}\psi$, 
$\tilde{g}_{ab}=\lambda^2g_{ab}$, will also satisfy the Weyl equation in 
the metric $\tilde{g}_{ab}$. In both coordinate systems we have employed 
so far, the retarded/advanced system $(u,v)$ and the Kruskal system $(U,V)$, 
the metric is conformal to a flat one of the form $du\,dv$ or 
$dU\,dV$. The Weyl equation for such flat metrics splits into a pair of 
decoupled equations for the positive/negative chirality spinors, 
$\psi_{\pm}$, which read
\begin{equation}
\partial_u\psi_+=\partial_v\psi_-=0 \hspace{1cm} (u,v)-\mbox{system},
\label{weyl}
\end{equation}
with identical equations in the $(U,V)$-system. In string terminology, 
$\psi_+$ is left moving and $\psi_-$ is right moving. The spinor in 
spacetime can now be obtained multiplying the flat $\psi_{\pm}$ with 
the respective conformal factors.\\ 
Normalizable, positive frequency solutions of (\ref{weyl}) in the interval 
$r_*\in(-\infty,0)$ are given by
\begin{equation}
\psi_+=\frac{e^{-i\omega v}}{\sqrt{2\pi}}, \hspace{1cm} 
\psi_-=\frac{e^{-i\omega u}}{\sqrt{2\pi}}.
\end{equation}
There is no question to impose Dirichlet or other boundary 
conditions here, because the Weyl equations fix a unique form to the 
positive frequency modes. Indeed, that Dirichlet boundary conditions on the 
Dirac equation lead to inconsistencies has long been known, as 
a consequence of the first order character of the equation. Note 
however that for the given solutions, the component of the conserved 
Dirac current along the normal to the boundary at infinity, namely 
$j=\bar{\psi}\rho^1\psi$ evaluated at $r_*=0$, vanishes identically. 
This is a much more weaker boundary condition than Dirichlet or 
Neumann, allowing to find non--trivial solutions. In the Kruskal frame, 
we take advantage from the fact that both horizons are Cauchy surfaces 
for Weyl spinors in anti-de Sitter space. The normalization measure is 
$(\kappa U)^{-1}dU$ along the past sheet of the horizon, and 
$(\kappa V^{-1})dV$ along the future sheet, so normalizable solutions are
\begin{equation}
\psi_+=\frac{(\kappa|V|)^{1/2}e^{-i\omega V}}{\sqrt{2\pi}} \hspace{1cm}
\psi_-=\frac{(\kappa|U|)^{1/2}e^{-i\omega U}}{\sqrt{2\pi}}.
\end{equation}
We shall use these to compute the relevant Bogoljubov coefficients in 
the next but one section. 

\subsection{Stress Tensor}

We now wish to determine the expectation value of the stress tensor 
\begin{equation}
T_{ab} = \phi,_a \phi,_b - \frac{1}{2}g_{ab}\phi,_c \phi^{,c} \label{stress}
\end{equation}
for the Unruh--, the Hartle--Hawking--, and the Boulware vacuum state 
(see \cite{wipf98} for a readable account).
Now, it is well known \cite{wald2} that (\ref{stress}) is mathematically
ill--defined, involving products of two distributions on spacetime.
Therefore some kind of regularization procedure is needed, e.~g.~a
point--splitting method.
We follow here the lines of Davies and Fulling \cite{davful}, 
who presented in
detail the renormalization theory of the stress tensor of a two--dimensional
massless scalar field, including boundary conditions.
They consider the line element
\begin{equation}
ds^2 = c(u,v)dudv,
\end{equation}
where $c(u,v)$ is the conformal factor. For mode functions of the form
\begin{equation}
\phi_{\omega} = (4\pi\omega)^{-\frac{1}{2}}(e^{-i\omega u} \pm e^{-i\omega v})
\end{equation}
the renormalized vacuum expectation value of the energy--momentum tensor
is given by
\begin{equation}
\langle T_{ab}\rangle = \theta_{ab} - (48\pi)^{-1}R g_{ab}, \label{stressren}
\end{equation}
where $R$ is the scalar curvature of the manifold, which gives
rise to the conformal anomaly, and $\theta_{ab}$
depends on the conformal factor $c$ according to
\begin{eqnarray}
\theta_{aa} &=& -(12\pi)^{-1}c^{-2}\left[\frac{3}{4}(\partial_a c)^2 -
                \frac{1}{2}
                c\partial^2_ac\right] \quad (a = u,v) \nonumber \\ 
\theta_{uv} = \theta_{vu} &=& 0.
\end{eqnarray}
In the following we omit the brackets for the expectation value of $T_{ab}$.
Using (\ref{stressren}), one obtains for the renormalized stress tensor
in the Boulware state
\begin{eqnarray}
T^B_{uv} &=& \frac{RV(r)}{96\pi} \nonumber \\ 
T^B_{uu} = T^B_{vv} &=& \frac{\eta}{16\pi r}\left(\frac{\eta}{r^3} -
                        \frac{2}{\ell^2}\right),
\end{eqnarray}
where $R = - d^2V/dr^2$.\\ 
In order to examine whether $T_{ab}$ is well--defined everywhere or not,
we have to express it in the global nonsingular coordinate system $(U,V)$.
The components then read
\begin{eqnarray}
T^B_{UU} &=& \frac{1}{\kappa^2 U^2}T^B_{uu} \nonumber \\ 
T^B_{VV} &=& \frac{1}{\kappa^2 V^2}T^B_{vv} \nonumber \\ 
T^B_{UV} &=& -\frac{RV(r)}{96\pi UV}.
\end{eqnarray}
Thus $T^B_{UU}$ diverges on the future horizon, whereas $T^B_{VV}$
diverges on the
past horizon.\\ 
In coordinates $r,t$ we have
\begin{eqnarray}
T_t^{B\,t} &=& -\frac{2}{V(r)}(T^B_{uu} + T^B_{uv}) \nonumber \\ 
T_r^{B\,r} &=& \frac{2}{V(r)}(T^B_{uu} - T^B_{uv}) \nonumber \\ 
T_r^{B\,t} &=& T_t^{Br} = 0,
\end{eqnarray}
which behaves on the horizon as
\begin{equation}
T_a^{Bb}|_{r \approx r_+} = -\frac{\eta}{16\pi r_+^2(r - r_+)}
\left(\begin{array}{cc} -1 & 0 \\ 
                         0 & 1 \end{array} \right).
\end{equation}
This behaviour is similar to that of the stress tensor in the Boulware state
for the Schwarzschild spacetime \cite{cand80}. There it diverges like
$(r - r_+)^{-2}$ in four dimensions, our exponent $-1$ instead of $-2$
stems from the reduction to two dimensions.\\ 
The energy density $\rho$ in the Boulware state, measured by a Killing
observer with
four--velocity $v^a = V(r)^{-1/2}\partial_t$ (i.~e.~at constant $r$), is 
given for large $r$ by
\begin{equation}
\rho = T^B_{ab}v^av^b = V(r)^{-1}T^B_{tt} = -\frac{r^2}{24\pi V(r)\ell^4}\simeq
-\frac{1}{24\pi\ell^2}.
\end{equation}
This represents the Casimir energy of the state, resulting from the boundary
conditions at infinity. The "would be" divergence for $r \to \infty$ 
(i.~e.~on the boundary) is a well--known behaviour of the Casimir 
energy, but here it is cancelled by the growing of the lapse at infinity. 
Later we shall see that in the HH state there is
an additional term in the expression for $\rho$, coming from
black hole radiation.\\ 
Let us now focus our attention on the Hartle--Hawking state.
First of all, note that the formulas given in \cite{davful} for
the renormalized stress tensor are only valid for time--independent
boundary conditions with respect to $T := \frac{1}{2}(V+U)$. This is not
the case here. Therefore we must introduce new coordinates $\bar{V} = V$,
$\bar{U} = -1/U$. In these coordinates the boundary $UV = -1$ is located
at $\bar{U} = \bar{V}$, i.~e.~at $\bar{X} := \frac{1}{2}(\bar{V}-\bar{U}) = 0$.
Reexpressing $T^{HH}_{ab}$ in Kruskal coordinates after the calculation,
one gets
\begin{eqnarray}
T^{HH}_{VV} &=& -\frac{1}{48\pi V^2}w(r) \nonumber \\ 
T^{HH}_{UU} &=& -\frac{1}{48\pi U^2}w(r) \nonumber \\ 
T^{HH}_{UV} &=& -\frac{Rg_{UV}}{48\pi},
\end{eqnarray}
where we have defined
\begin{equation}
w(r) := -1 + \frac{V'(r)^2}{4\kappa^2} - \frac{V(r)V''(r)}{2\kappa^2}.
\label{w}
\end{equation}
Now near the horizon we have $w(r) \approx - 6e^{\pi/\sqrt{3}} U^2V^2$,
cancelling the divergence of the prefactor,
therefore $T^{HH}$ is well--defined everywhere.\\
Here a new feature arises in the higher genus case for $\eta<0$,
i.~e.~when an inner horizon forms. Using the $g>1$ expression for $V(r)$
in (\ref{w}), and transforming the stress tensor to coordinates in which
the metric is regular at the inner horizon $r_-$, one finds a divergence
of the stress tensor at $r=r_-$. Having adjusted the quantum state of 
the scalar field so that the stress tensor is finite at the outer 
horizon, it diverges at the inner horizon. This behaviour is identical 
to what has been found in two dimensions for a Reissner--Nordstr\"om 
black hole \cite{triv93}, and is related to an infinite frequency blue-shift 
occurring on the Cauchy horizon. This suggests that there is no
Hartle--Hawking equilibrium state, and that strong back
reaction effects take place near the inner horizon. In two dimensions, 
however, a semiclassical calculation shows that this last sentence is not 
true, in the sense that the singularity appearing at the horizon is very 
mild \cite{triv93}.\\
Let us now return to the toroidal black hole.
We observe that the $V$--component of the energy current four--vector $J^a$,
measured by an "observer" with four--velocity $v^a = \partial_U$ on the
past--horizon, namely
\begin{equation}
J^V = - T^{HH\,V}_b v^b = - g^{VU} T^{HH}_{UU}
\end{equation}
vanishes on $H_-$. Therefore $|H\rangle$ does not contain Kruskal particles
emerging from the white hole, as already pointed out in the previous
section.\\ 
For large $r$, the energy density $\rho$, measured by a Killing
observer with four--velocity $v^a=V(r)^{-1/2}\partial_t$, is now given by
\begin{equation}
\rho = V(r)^{-1}T_{tt}^{HH} = \frac{\kappa^2}{24\pi V(r)} - 
\frac{r^2}{24\pi V(r)\ell^4}.
\end{equation}
Again we meet the Casimir energy $-1/24\pi\ell^2$, but now there is
an additional term. It is connected to the thermal radiance of the
black hole, as can be readily seen by calculating the radiated energy
\begin{equation}
E = \frac{1}{2\pi}\cdot 2 \cdot \int_0^{\infty}\frac{\omega d\omega}
    {e^{2\pi\omega/\kappa} - 1},
\end{equation}
which yields exactly $\kappa^2/24\pi$. ($1/2\pi$ is the density of states
in one spatial dimension, and the additional factor 2 has to be included,
because we have left-- and right--moving waves). The factor 
$V(r)^{-1}$ is the usual Tolman red--shift of the local temperature, 
making the energy density vanishingly small at infinity.\\ 
We now want to show that $|H\rangle$ describes a thermal equilibrium state.
To this aim, we calculate the net null flux through a surface $r$ = const.
The energy current
four--vector $J^a$, measured by an observer at
constant $r$ with four--velocity $v^a$, is represented by
\begin{equation}
J_a = -T^{HH}_{ab}v^b.
\end{equation}
Now the net energy flux through the surface $r$ = const.~, i.~e.~$UV$ =
const.~$<0$, is given by the integral
\begin{equation}
\int_{UV = const.} J_a n^a dS,
\end{equation}
where $n^a$ denotes the unit normal to the surface, and $dS$ is the
induced "volume" element.
One easily verifies that for $UV$ = const.~the integrand $J_a n^a$ is zero,
hence the incoming null flux through the constant $r$ surface equals the
outgoing, and
$|H\rangle$ represents indeed a thermal equilibrium state with the Hawking
temperature.\\ 
For the Unruh vacuum we find a stress tensor identical to that of the
Hartle--Hawking state. This suggests that in our case these two states
coincide. Indeed, the Unruh modes (\ref{Umodes}) can be obtained from
the HH modes (\ref{HHmodes}) by interchanging $U$ and $V$.
This is an isometry
of the metric (\ref{metri}), because it leaves $r$ invariant and maps
$t$ to $-t$. Due to the reflective boundary conditions we were compelled
to impose (for all three choices in (\ref{bound}) the component
of the Klein--Gordon current normal to the boundary at infinity vanishes),
no Unruh state like in the Schwarzschild case can be defined, as all the
Hawking radiation emitted by the black hole is reflected at infinity
and travels back to the future horizon. Therefore necessarily a
thermal equilibrium state results, and $|U\rangle$ coincides
with $|H\rangle$.

\subsection{Black Hole Temperature and Luminosity}

First of all, let us remark that in principle the definition of a luminosity
makes only sense if there exists an Unruh state. As in the case under
consideration the Unruh state also describes thermal equilibrium,
i.~e.~the black hole absorbs the same amount of radiation as it emits,
the net luminosity is zero. Nevertheless we can calculate the emitted
radiation (which, of course, is reflected at the boundary), and call this
the luminosity of the black hole.
To this end,
we have to find the Bogoljubov transformation relating Unruh modes to
Boulware modes, i.~e.
\begin{equation}
\varphi_{\omega}^B(u,v) = \int_0^{\infty}(\alpha_{\omega \omega'}
\varphi_{\omega'}^{U}(U,V) + \beta_{\omega \omega'}
\bar{\varphi_{\omega'}}^{U}(U,V))d\omega'. \label{bogol}
\end{equation}
A nice calculation yields
\begin{eqnarray}
\alpha_{\omega \omega'} &=& -\frac{i}{2\pi}\frac{\omega'^{-i\omega/\kappa}}
{\omega \omega'}\Gamma\left(1 + i\frac{\omega}{\kappa}\right)\exp\left(
\frac{\pi\omega}{2\kappa}\right), \\ 
\beta_{\omega \omega'} &=& \frac{i}{2\pi}\frac{\omega'^{-i\omega/\kappa}}
{\omega \omega'}\Gamma\left(1 + i\frac{\omega}{\kappa}\right)\exp\left(
-\frac{\pi\omega}{2\kappa}\right).
\end{eqnarray}
Inserting (\ref{bogol}) into equation (\ref{expanB}), we obtain the
relation between Boulware and Unruh operators
\begin{eqnarray}
b_{\omega'} &=& \int_0^{\infty} d\omega [\bar{\alpha}_{\omega' \omega}
                a_{\omega} - \bar{\beta}_{\omega' \omega}a_{\omega}^{\dagger}],
\end{eqnarray}
The matrix element $\langle U\,|\,b_{\omega}^{\dagger}
b_{\omega'}\,|\,U\rangle$ can now be
calculated, and we obtain
\begin{equation}
\langle U\,|\,b_{\omega}^{\dagger}b_{\omega'}\,|\,U\rangle =
\int_0^{\infty}\beta_{\omega k}
\bar{\beta}_{\omega' k}dk = \frac{\delta(\omega - \omega')}{\exp\left(
\frac{2\pi\omega}{\kappa}\right) - 1}.
\end{equation}
Since $\delta(0)=T/2\pi$ for very large time $T$, this means
that on the average there are
\begin{equation}
dn_{\omega} = \frac{1}{2\pi}\frac{d\omega}{\exp
\left(\frac{2\pi\omega}{\kappa}\right) - 1}
\label{flux}
\end{equation}
zero rest mass particles flowing near infinity per unit time
in the frequency range
between $\omega$ and
$\omega + d\omega$. As already pointed out, the radiation is reflected
at infinity, so the net particle flux is zero. 
From (\ref{flux}) we also
infer that the radiation temperature of the black hole is $T = \kappa/2\pi$,
in accordance with the temperature found in the Euclidean formulation
\cite{vanz97,bril97}.\\ 
Note that, in order to obtain the "luminosity", it would have been sufficient 
to calculate the Bogoljubov coefficients relating an outgoing mode
$\exp(-i\omega U)$ to the modes $\exp(-i\omega u)$.
This yields the same Bogoljubov coefficients as above, which can be
understood as follows:
The monochromatic components $\exp(\pm i\omega u)$ in the expansion of the
outgoing mode $\exp(-i\omega U)$ are reflected at the boundary, becoming
ingoing modes $\exp(\pm i\omega v)$. These give the expansion of
the reflected wave
in Kruskal coordinates, namely of $\exp(i\omega/V)$.\\ 
At this point one may ask how it is possible to assign the outgoing
flux only to the modes $\exp(-i\omega U)$, and the reflected one only
to $\exp(i\omega/V)$. This is a legitimate question, in view of the
stress tensor being quadratic in $\phi$. However, the energy current
four--vector $J$ measured by an observer with
four--velocity $V(r)^{-1/2}\partial_t$, is given by
\begin{eqnarray}
J^u &=& -g^{uv}V(r)^{-1/2}(T_{uv} + T_{vv}) \nonumber \\ 
J^v &=& -g^{uv}V(r)^{-1/2}(T_{uv} + T_{uu}),
\end{eqnarray}
and one can show that
only the modes $\exp(-i\omega U)$ contribute to $T_{uu}$, and only
the $\exp(i\omega/V)$
contribute to $T_{vv}$ \cite{davful}. 
As $T_{uv}$ is completely fixed by the manifold
via the conformal anomaly, it is independent of the modes. Therefore
the outgoing null flux $J^v$ is determined by the modes $\exp(-i\omega U)$
only, whereas the reflected (ingoing) flux is determined exclusively by the
$\exp(-i\omega/V)$.\\ 
For the Weyl fermions the relevant Bogoljubov coefficient is
\begin{equation}
\beta_{\omega \omega'} = e^{i\pi/4}\frac{1}{2\pi\sqrt{\kappa\omega'}}
                         e^{-\pi\omega/\kappa}
                         \omega'^{-i\omega/\kappa}\Gamma
                         \left(\frac{1}{2}+\frac{i\omega}
                         {\kappa}\right),
\end{equation}
from which one obtains the Fermi--Dirac distribution
\begin{equation}
\int_0^{\infty}\bar{\beta}_{\omega\sigma}\beta_{\omega'\sigma}\,d\sigma =
\frac{\delta(\omega-\omega')}{e^{2\pi\omega/\kappa}+1}.
\end{equation}

\section{Generalization to Four Dimensions} \label{4d}

We now consider the full metric (\ref{metri}) and recall some results
about the family of black holes it describes. We will not specify a 
particular horizon metric for $g>1$ (which depends on the $6g-6$ 
moduli of a Riemann surface), since its precise form is not important 
in what follows. For $g=1$ the moduli space of the torus 
is ${\rm H}^+/{\rm SL}(2,Z)$ (${\rm H}^+$ denotes the
upper complex half plane), and the torus flat metric is given by 
Eq.~(\ref{tmet}) in terms of its Teichm\"uller parameter $\tau\in 
{\rm H}^+$.  Any two
such parameters related by ${\rm SL}(2,Z)$ fractional linear 
transformations describe conformally equivalent tori. \\ 
The metric (\ref{metri}) can be continued to imaginary values of the 
Killing time (${\cal T} = it$) as a Riemannian non--singular
metric everywhere. This 
metric takes the form 
\begin{equation}
ds^2=V(r)d{\cal T}^2+V^{-1}(r)dr^2+r^2d\sigma^2,
\label{Emetri}
\end{equation}
where we recall that $\Lambda=-3/\ell^2$ and
\begin{eqnarray}
V(r) = -1 + \delta_{g,1} -\frac{2\eta}{r} + \frac{r^2}{\ell^2},
\end{eqnarray}
and $r>r_+$ is required for this to be of positive signature. 
Close to the horizon $V(r)=2\kappa(r-r_+)$, where $\kappa$ is the surface 
gravity, and the near horizon geometry is described in terms of 
proper distance, $s^2=2(r-r_+)/\kappa$, by the metric
\begin{eqnarray}
ds^2=\kappa^2s^2d{\cal T}^2+ds^2+r_+^2d\sigma^2
\end{eqnarray}
Regularity of the metric then requires the period of ${\cal T}$ to be 
$\beta_+=2\pi\kappa^{-1}$, so this fixes the black hole's temperature. 
The zero temperature state has $\kappa=0$ and is a naked singularity
with parameter $\eta=0$ for $g=1$, and an extremal black hole
with parameter $\eta=-\ell/3\sqrt{3}$ for $g>1$,
while the positive temperature states 
above it have positive mass.
The member of the family with mass 
$M=\ell(g-1)/\sqrt{27}$ has parameter $\eta=0$ and is the quotient of 
anti--de Sitter space by a discrete subgroup of its isometry group 
\cite{amin97}, in particular it is a space of constant 
curvature. \\ 
Along with the metric, one can analytically continue the wave 
equation. This then gives an elliptic operator with non--singular 
coefficients and positive spectrum. The Schwinger function is the 
symmetric two--point function which decays to zero at infinity, it is 
regular at the origin and solves the Euclidean wave equation. As for 
the metric, regularity at the origin (the horizon in the Lorentzian 
sector) demands that the Schwinger function be periodic in ${\cal T}$ with 
period $\beta_+$. The analytically continued function in real time will 
then be periodic in imaginary time and regular all over the event 
horizon, but for $\eta<0$, the function can be extended only up to the 
inner Cauchy horizon. The quantum state to which it corresponds is
the equilibrium 
Israel--Hartle--Hawking state, and describes a topological black hole in 
thermal equilibrium with black body radiation at the Hawking
temperature. Later we shall discuss this state from a proper quantum 
field theory approach. \\ 
The contribution of the black hole to the partition function is $\ln 
Z=-I_E$, the on--shell value of the Euclidean action of the black 
hole \cite{gibb77}. 
The Euclidean action can also be evaluated off--shell ($\beta\neq\beta_+$), 
relative to the zero temperature ground state, and is \cite{vanz97}
\begin{eqnarray}
I_E=\beta M-\frac{A}{4},
\end{eqnarray}
where $A$ is the area of the event horizon. Since $M$ and $\beta$ 
are here independent variables, this quickly leads to an entropy 
$S=A/4$ which, when expressed as a function of the mass, has the large 
mass behaviour $S\simeq CM^{2/3}$. This means that the density of 
states grows as $\exp(CM^{2/3})$, so the partition function will 
converge. This is not a special feature of topological black holes, but 
also holds for the genus-$0$ anti--de Sitter black hole \cite{page83} 
and is related to a negative cosmological constant rather than to 
topology.
 
\subsection{The Israel--Hartle--Hawking State}

\begin{figure}[p]
\begin{center}
\unitlength1cm
\begin{picture}(10,10)
\epsfxsize15cm
\put(-2.5,0){\epsfbox{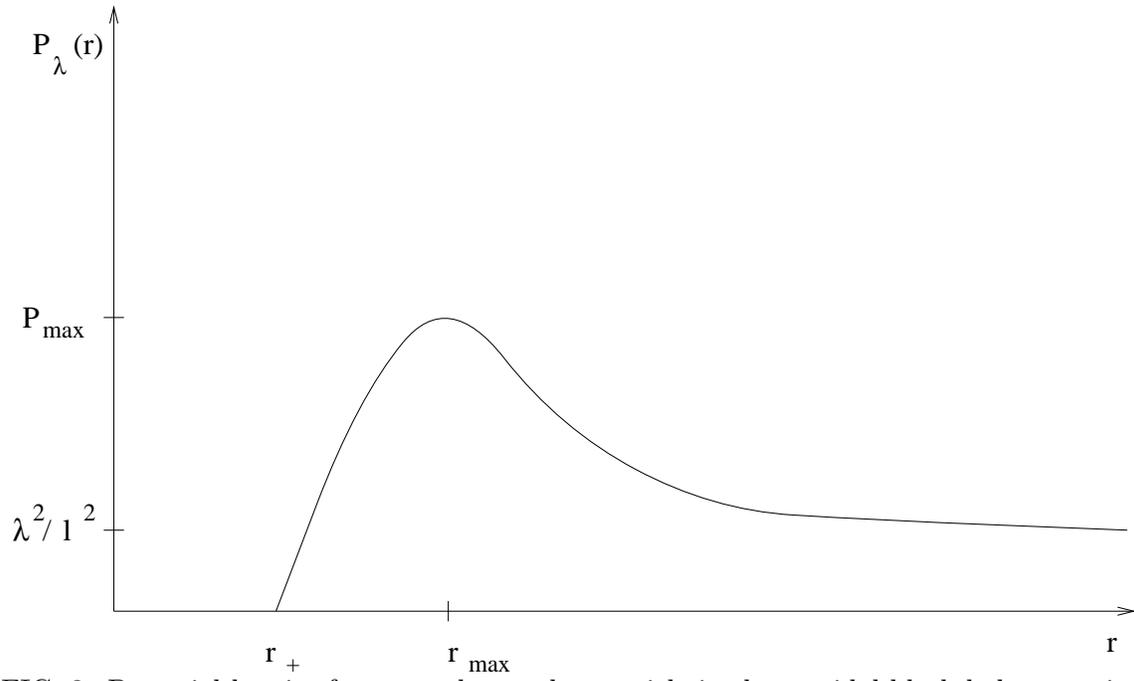}}
\end{picture}
\caption{Potential barrier for a massless scalar particle in the toroidal
black hole spacetime, or in the $g>1$ spacetime with $\eta>0$.
\label{pot}}
\end{center}
\end{figure}

\begin{figure}[p]
\begin{center}
\unitlength1cm
\begin{picture}(10,10)
\epsfxsize15cm
\put(-2.5,0){\epsfbox{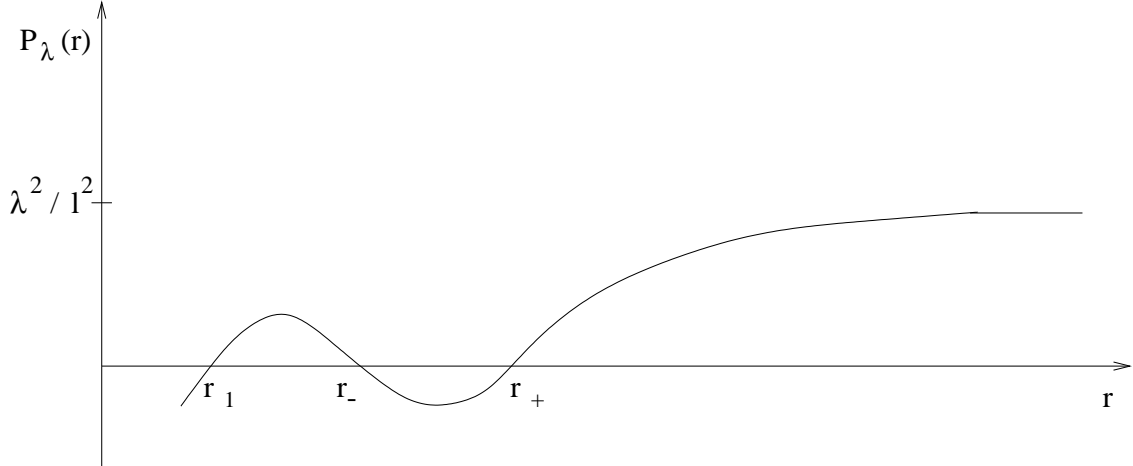}}
\end{picture}
\caption{Potential barrier for a massless scalar particle
in the $g>1$ spacetime with $\eta<0$. $r_1$ is the zero of the expression
$(\lambda^2/r^2+2\eta/r^3)$, appearing in the potential (\ref{barrier}).
In the figure the case $\lambda^2 > 2|\eta|/r_-$ is shown. For
$2|\eta|/r_+ < \lambda^2 < 2|\eta|/r_-$ one has $r_- < r_1 < r_+$,
and for $\lambda^2 < 2|\eta|/r_+$ one has $r_1 > r_+$.
(The course of the potential
is the same in all three cases).
\label{pot1}}
\end{center}
\end{figure}

To discuss black hole emission, we shall consider a scalar 
field obeying the conformally invariant Klein-Gordon equation
\begin{equation}
\frac{1}{\sqrt{-g}}\partial_a(g^{ab}\sqrt{-g}\partial_b)\phi -
\frac{1}{6}R \phi = 0.
\end{equation}
This equation can be separated into the following eigenvalue equations: 
Setting $\phi=r^{-1}F_j(t,r)u_j(x)$ and $\partial_*=\partial_{r_*}$, 
we have a two-dimensional wave equation 
for $F_j$, with a potential barrier $P_{\lambda}(r)$,
\begin{equation}
\partial_t^2F_j-\partial_{*}^2F+P_{\lambda}(r)F_j=0, \label{equF}
\end{equation}
together with the eigenvalue equation for the Laplacian on $S_g$
\begin{equation}
\Delta u_j=-\lambda_j^2u_j,
\label{eigen}
\end{equation}
where the potential barrier is given by
\begin{equation}
P_{\lambda}(r)=V(r)\left(\frac{\lambda^2}{r^2}+\frac{V^{'}(r)}{r}
+\frac{R}{6}\right)=V(r)\left(\frac{\lambda^2}{r^2}+\frac{2\eta}{r^3}
\right), \label{barrier}
\end{equation}
and the scalar curvature is $R = -12\ell^{-2}$. This term precisely 
cancels the divergent (as $\propto r^2$ at infinity) anti--de Sitter 
gravitational potential, which is why conformal scalar emission will 
be greater than that of minimally coupled scalars. For these the 
barrier is parabolic at infinity, with behaviour $P_{\lambda}(r)\simeq 
2\ell^{-4}r^2$, and the modes behave like Bessel functions at 
infinity.\\
In all cases the potential vanishes at the horizon and approaches
the asymptotic value $\lambda^2/\ell^2$ for $r \to \infty$.
For $g=1$ or $g \ge 1$ and $\eta > 0$ we have also a local
maximum outside the horizon.
The behaviour of $P_{\lambda}(r)$ is shown in figure \ref{pot} for the 
torus or a $g>1$ black hole with $\eta>0$, and in figure \ref{pot1} for 
$g>1$ and $\eta<0$.
Again, new features arise when $\eta<0$, i.~e.~when 
an inner horizon forms. For sufficiently large eigenvalues $\lambda^2$, 
there is a potential well in between the two horizons, causing 
amplification for waves entering from the outer horizon
(c.~f.~Chandrasehkar's monography \cite{chandra}). 
For $g>1$ and $\eta=0$, the potential is zero at the horizon, and
then monotonically increases to reach the asymptotic value
$\lambda^2/\ell^2$.\\
Let us consider now the eigenvalue equation (\ref{eigen}). 
On a general Riemann surface there are comparatively little
informations on the eigenvalues $\lambda_j$, except that they are finitely 
degenerate and form an unbounded increasing sequence. For the torus 
we have instead an exact formula for all the eigenvalues and, at least
for $\tau = i$ (symmetric torus), 
for the respective degeneracies. Indeed, solutions of (\ref{eigen}) 
must be automorphic functions under the identification group
\begin{eqnarray}
x & \simeq x+n &\hspace{1cm} n\in Z\\y & \simeq y+m &\hspace{1cm} m\in Z.
\end{eqnarray}
This fixes the normalized eigenfunctions to be (the torus area element 
is $dS = \mbox{Im}\tau dx\,dy$)
\begin{equation}
u_{nm}(x,y) = \sqrt{\mbox{Im}\tau}^{-1}\exp(2\pi i(nx+my)),
\end{equation}
and therefore the eigenvalues are  
\begin{equation}
\lambda_{nm}^2=(2\pi)^2n^2+
\left(\frac{2\pi}{\mbox{Im}\tau}\right)^2\,(m-n\mbox{Re}\tau)^2.
\end{equation}
For arbitrary values of the Teichm\"uller parameter $\tau$ the
degeneracy $g_{\lambda}$ is difficult (if not impossible) to calculate.
For $\tau = i$ however,
$g_{\lambda}$ equals the number of representations of 
$\lambda^2/4\pi^2$ in the form 
\begin{equation}
\frac{\lambda_{nm}^2}{4\pi^2} = n^2+m^2,
\end{equation}
which is given by \cite{hardy}
\begin{equation}
g_{\lambda} = 4 \sum_{d|\frac{\lambda^2}{4\pi^2}}\chi(d),
\end{equation}
where one has to sum over all divisors $d$ of $\lambda^2/4\pi^2$, and
$\chi(d)$ is defined by
\begin{equation}
\chi(d) = \left\{ \begin{array}{c@{\quad,\quad}c}
          0 & 2 \mid d \\ 
          (-1)^{\frac{1}{2}(d-1)} &
          2 {\hbox{{$\,\,\mid$}\kern-.5em\hbox{$\not$\quad}}} d.
          \end{array} \right.
\end{equation}
Knowing the eigenvalues, we can write explicitly 
the potential barrier felt by a mode in a toroidal black hole, which is
\begin{equation}
P_{\lambda}(r)=\left(\frac{r^2}{\ell^2}-\frac{2\eta}{r}\right)\left[
\frac{\lambda_{nm}^2}{r^2}+\frac{2\eta}{r^3}\right].
\end{equation}
The potential is zero at the horizon. (It falls off exponentially
in the tortoise coordinate $r_*$ for $r_* \to -\infty$, i.~e.~on the
horizon). There is a maximum of $P_{\lambda}(r)$ at $r=r_{max}$, where
\begin{equation}
r_{max} = \left\{ \begin{array}{c@{\quad,\quad}c@{\qquad}c}
          2\lambda_{mn}\ell\cosh\frac{\varphi}{3} & \lambda_{mn}^3 <
          \frac{4\eta}
          {\ell} & (\cosh\varphi := \frac{4\eta}{\ell\lambda_{mn}^3}) \\ 
          2\lambda_{mn}\ell\cos\frac{\varphi}{3} & \lambda_{mn}^3 \ge
          \frac{4\eta}
          {\ell} & (\cos\varphi := \frac{4\eta}{\ell\lambda_{mn}^3}).
          \end{array} \right.
\end{equation}
For $\lambda=0$, $r_{max}=4^{1/3}r_+$ is just outside the black hole, 
and it is increasing to infinity as $\lambda\to\infty$.
For $\lambda_{mn}^3 < 4\eta/\ell$ the potential maximum
$P_{\lambda}(r_{max})$ is given by
\begin{equation}
P_{\lambda}(r_{max}) = \frac{3\left(\frac{\eta}{\ell} + \lambda_{mn}^3
\cosh\frac{\varphi}{3}\right)^2}{4\lambda_{mn}^4\ell^2\cosh^4\frac{\varphi}
{3}}. \label{Pmax}
\end{equation}
For $\lambda_{mn}^3 \ge 4\eta/\ell$ the $\cosh$ has to be replaced by
a $\cos$. At infinity the potential equals
the constant $\lambda_{nm}^2\ell^{-2}$. 
The potential curve is shown in figure \ref{pot}.\\ 
For any genus, a set of Boulware modes, normalized to $\delta(\omega)$, 
can be defined by the asymptotic conditions
\begin{eqnarray}
\lefteqn{B_{\omega\lambda}\simeq} \nonumber \\ 
&&(4\pi\omega)^{-1/2}e^{-i\omega t}u_{\lambda}(x,y)
r^{-1}\left\{\begin{array}{cc}e^{i\omega 
r_*} + R_{\lambda}(\omega)e^{-i\omega r_*} & r_*\to-\infty \\ 
T_{\lambda}(\omega)\sin(\Omega_{\lambda} r_*) & r_*\to 0,
\end{array}\right.  \label{boul}
\end{eqnarray}
where $u_{\lambda}(x,y)$ are the eigenfunction of the scalar Laplacian 
on a Riemann surface of genus $g \ge 1$, 
$\Omega_{\lambda}=\sqrt{\omega^2-\lambda_{nm}^2\ell^{-2}}$, and 
$R_{\lambda}(\omega)$ and $T_{\lambda}(\omega)$ are the reflection--
and the transmission coefficients of the potential barrier, 
respectively.\\ 
For $\Omega_{\lambda}$ 
imaginary, the mode at infinity acquires an additional phase $\pm i$. 
This is a Dirichlet set, but a Neumann set can also be defined by 
replacing the sine function with a cosine. The modes appear to 
emerge from the past horizon in the eternal black hole spacetime.\\
The phase of $R_{\lambda}$ is then twice the phase of $T_{\lambda}$, as a 
consequence of the boundary condition. These phase shifts have no 
singularities in the lower half complex $\omega$--plane, because the 
potential admits no bound states. Note that for $\eta<0$, the 
potential for modes with sufficiently small eigenvalues has a well 
outside the event horizon (see figure \ref{pot1}). This occurs for 
$0<\lambda^2<2/3$, and such small eigenvalues exist in general on any 
Riemann surface. The classical counterpart is that there are no closed 
null geodesics around the black hole within the potential well. 
Resonant diffusion is not excluded, but we have not analyzed this any 
further (at the large frequencies which are relevant to the Hawking 
radiation, there is certainly no problem with resonances).\\
The Dirichlet coefficients $R_{\lambda}(\omega)$ and $T_{\lambda}(\omega)$
are not related by current conservation, due to the boundary conditions. 
However, they can be related to the coefficients describing scattering 
off the barrier without the boundary conditions at infinity, i.~e.~by
replacing $\sin\Omega_{\lambda}r_*$ by $\exp(\pm i\Omega_{\lambda}r_*)$ in
(\ref{boul}). We shall denote these outgoing 
reflection/transmission coefficients by right pointing arrows, 
$\stackrel{\rightarrow}{R}_{\lambda}\!(\omega)$ and 
$\stackrel{\rightarrow}{T}_{\lambda}\!(\omega)$ 
respectively, and the ingoing coefficients with left pointing arrows.
Current conservation then gives the unitarity conditions
\begin{eqnarray}
\omega[1-|\!\stackrel{\rightarrow}{R}_{\lambda}(\omega)|^2]=
\sqrt{\omega^2-\lambda^2
\ell^{-2}}\,|\!\stackrel{\rightarrow}{T}_{\lambda}\!(\omega)|^2
\label{unit}
\end{eqnarray}
\begin{eqnarray}
\sqrt{\omega^2-\lambda^2\ell^{-2}}\stackrel{\rightarrow}{T}_{\lambda}\!(\omega)
=\omega\stackrel{\leftarrow}{T}_{\lambda}\!(\omega).
\end{eqnarray}
The original coefficients are then given in terms of 
$\stackrel{\rightarrow}{R}_{\lambda}\!(\omega)$ and
$\stackrel{\rightarrow}{T}_{\lambda}\!(\omega)$ 
by the equations
\begin{eqnarray}
R_{\lambda}(\omega)=-\frac{Z}{\bar{Z}},\qquad T_{\lambda}(\omega)=
-\frac{2i|\!\stackrel{\rightarrow}{T}_{\lambda}\!(\omega)|^2}{\bar{Z}},
\label{relat}
\end{eqnarray}
where 
$Z=\stackrel{\rightarrow}{R}_{\lambda}\!(\omega)\stackrel{\rightarrow}
{\bar{T}}_{\lambda}\!(\omega) -
\stackrel{\rightarrow}{T}_{\lambda}\!(\omega)$.\\ 
Clearly, with either boundary 
conditions $R_{\lambda}(\omega)=\exp(i\delta_{\lambda}(\omega))$, so 
all the emitted radiation is ultimately reflected back 
into the black hole. In fact an eternal black hole can only
exist in a thermal equilibrium state.\\ 
To introduce this equilibrium state, we will find the solutions of the 
wave equation that are positive frequency along one sheet of the event 
horizon, with respect to its canonical affine parameter.\\ 
We define the Hartle--Hawking modes to be solutions of the wave 
equation which obey Dirichlet boundary conditions at infinity and are positive 
frequency on the past horizon $H^-$, with respect to the 
canonical affine parameter $U$, i.e. $\partial_UH_{\lambda\omega}=
-i\omega H_{\lambda\omega}$. 
Outside the horizon they will be superpositions of Boulware 
modes, which we write in the form
\begin{eqnarray}
H_{\omega\lambda}(p)=\int_0^{\infty}\frac{d\omega'}{\sqrt{4\pi\omega'}}
[\bar{\gamma}_{\omega'\omega}
B_{\omega'\lambda}(p)-\epsilon_{\omega'\omega}\bar{B}_{\omega'\lambda}(p)],
\label{IHH}
\end{eqnarray}
where $p=(u,v,x,y)$ belongs to the outer region. The boundary conditions
at infinity are then
automatically satisfied. 
By definition, on the past horizon $H_{\lambda\omega}(p)$ converges
to the function 
$(4\pi\omega)^{-1/2}\exp(-i\omega U)u_{\lambda}$. Using the Fourier transform
\begin{eqnarray}
e^{-i\omega U}\theta(-U) = \int_{-\infty}^{\infty}\frac{d\omega'}{2\pi\kappa}
                 e^{\pi\omega'/2\kappa}\omega^{i\omega'/\kappa}
                 \Gamma(-i\omega'/\kappa)e^{-i\omega'u} 
\label{trans}
\end{eqnarray}
and the asymptotic condition (\ref{boul}), we find that the phase 
$\delta_{\lambda}(\omega)$ of $R_{\lambda}(\omega)$ disappears along $H^-$ 
and we get the $\gamma$- and $\varepsilon$-coefficients in the form
\begin{eqnarray}
\gamma_{\omega'\omega} &=& -\frac{i}{2\pi}\frac{\omega^{-i\omega'/\kappa}}
                                {\sqrt{\omega\omega'}}\Gamma(1+i\kappa^{-1}
                                \omega')e^{\pi\omega'/2\kappa} \nonumber \\ 
\epsilon_{\omega'\omega} &=&\frac{i}{2\pi}\frac{\omega^{-i\omega'/\kappa}}
                                {\sqrt{\omega\omega'}}\Gamma(1+i\kappa^{-1}
                                \omega')e^{-\pi\omega'/2\kappa}.
\label{coeff}
\end{eqnarray}
On the future horizon $H^+$, we now claim that $H_{\omega\lambda}(p)$ is a 
superposition of positive frequency solutions with respect 
to the canonical affine parameter 
$V$ of the future horizon. In fact, using (\ref{boul}) and going 
on $H^+$ we obtain
\begin{eqnarray}
H_{\omega\lambda}(p)=\int_0^{\infty}\frac{d\omega'}{\sqrt{4\pi\omega'}}
[\bar{\gamma}_{\omega'\omega}
e^{-i\omega'v+i\delta_{\lambda}(\omega')}-\epsilon_{\omega'\omega}
e^{i\omega'v-i\delta_{\lambda}(\omega')}], 
\label{IHH+}
\end{eqnarray}
where now the phase shifts do give a contribution to the mode.\\ 
From Eq.~(\ref{boul}) and the uniqueness of Dirichlet solution 
we deduce $B_{-\omega\lambda}=i\bar{B}_{\omega\lambda}$, up to a 
phase, so 
$R_{\lambda}(-\omega)=\bar{R}_{\lambda}(\omega)$, which gives the antisymmetry 
condition $\delta_{\lambda}(-\omega)=-\delta_{\lambda}(\omega)$.\\ 
We are now in position to prove our claim. One may recast (\ref{IHH+}) 
in the form of a single integral over the real line, as the pole in 
$\omega'=0$ in the two integrals cancel. 
From Eq.~(\ref{coeff}), the functions $\gamma_{\omega'\omega}$ and
$\varepsilon_{\omega'\omega}$ 
have infinitely many 
simple poles in the lower half complex $\omega'$--plane, at
$\omega'_n=-i\kappa n$
for $n\geq0$ and integer. As the potential barrier has no bound 
states, the phase shifts $\delta_{\lambda}(\omega')$ have no poles for 
$\mathop{\rm Im}\nolimits\omega'<0$. By analytic continuation arguments,
the Boulware modes 
along the imaginary axis are real functions, so (\ref{relat}) gives 
$R_{\lambda}(-i\omega)=-1$, or $\delta_{\lambda}(-i\omega)=\pm\pi$.
We may then compute 
$H_{\lambda\omega}$ along $H^+$ by summing over the residues (omitting 
the pole in $0$) and putting 
$\delta_{\lambda}(-in\kappa)=\pm\pi$, after which we obtain the result
\begin{eqnarray}
H_{\lambda\omega}(p)=(4\pi\omega)^{-1/2}\left[1-
\exp\left(\frac{i\omega}{V}\right)\theta(V)\right], \qquad p\in H^+.
\end{eqnarray} 
The "$1$" above is the relict of $\exp(-i\omega U)$ along $H^+$ (which 
is the set $U=0$) and, apart from it, 
the function $H_{\lambda\omega}(p)$ is analytic for $\mathop{\rm Im}
\nolimits V<0$, so 
its Fourier expansion must contain only 
positive frequencies (it is actually a superposition of Bessel functions of 
order zero).\\ 
We have obtained an interesting result. We started with a function 
like $\exp(-i\omega U)$ along $H^-$, as in the definition of the Unruh 
vacuum, and we ended with a function like $\exp(i\omega V^{-1})$ along 
$H^+$, which is a superposition of positive frequency $V$-modes. This 
means that the state defined by the modes $H_{\lambda\omega}$ is a true vacuum 
for particles defined in the Kruskal time, $T=(U+V)/2$, and therefore 
in particular it is an equilibrium state. This is the 
Israel--Hartle--Hawking state $|H\rangle$: writing the quantum field as
\begin{eqnarray}
\phi(p)=\int_0^{\infty}[A_{\lambda\omega}H_{\lambda\omega}(p)+
A_{\lambda\omega}^{\dag}
\bar{H}_{\lambda\omega}(p)]\,d\omega,
\end{eqnarray}
it is defined by $A_{\lambda\omega}|H\rangle = 0$. When analyzed in terms of 
Boulware modes, however, we will find it to contain a thermal 
distribution of particles with the black hole's temperature.\\ 
From the above it also follows that we could have defined 
the Hartle--Hawking 
modes to be positive frequency along the future horizon with respect 
to $V$ (which would be the usual definition for asymptotically flat black 
holes). In this case we would have ended with modes which are positive 
frequency along the past horizon in the time $U$, and therefore we would
not have 
changed the definition of the state.\\ 
We see then that there is no Unruh state, since 
modes which are positive frequency relative to $U$ along $H^-$ 
and obey Dirichlet boundary conditions at infinity, are also positive 
frequency along $H^+$ in the time $V$.\\ 
For $\eta>0$, the Hartle--Hawking modes are defined  everywhere. 
For $\eta<0$, the Hartle--Hawking modes are defined in the region 
contained within the inner Cauchy horizon, where they stay bounded. 
However, any flux of energy coming from outside the event horizon 
diverges relative to a local frame crossing the Cauchy horizon 
\cite{chandra}, due to an infinite blue shift. Hence we suspect that 
there will be divergences in the quantum expectation value of the 
stress tensor in the Hartle--Hawking state, near the Cauchy horizon (in 
two dimensions it diverges, in fact). If this is the case, then one 
cannot ignore the back raction of the thermal energy on the spacetime, 
as it is done implicitly in defining the Hartle-Hawking state. In the 
analogous situation of a Reissner--Nordstr\"{o}m black hole, this 
question is yet unsettled to the authors' knowledge (c.~f.~\cite{hisc} 
for this case), since the Killing approximation devised by 
Frolov--Zel'nikov \cite{frol} fails near the horizons, as well as the analytic 
approximation devised by Hiscock et al.~\cite{hisc}. However, this is 
a global question that will not affect our subsequent results.\\
From Eq.~(\ref{coeff}) and the expansion (\ref{IHH}), we can easily 
determine the mean occupation number near infinity, for Boulware 
particles with energy $\omega$ 
(i.~e.~for particles defined by the static time parameter) in the 
Israel--Hartle--Hawking state: it is a Planck distribution with the 
black hole's temperature
\begin{equation}
dN_{\lambda\omega} = \frac{g_{\lambda}\,d\omega}{e^{2\pi\omega/\kappa}-1}, 
\label{numbpart}
\end{equation}
where $g_{\lambda}$ is the degeneracy of $\lambda$. To find the energy density 
from this is slightly non--trivial, as one would use the density of states 
and then sum over the degeneracy $g_{\lambda}$. We will present a 
calculation of this kind when discussing particle production by the 
black hole, using Weyl's asymptotic formula.\\ 

\subsection{Particle Production}

In a thermal equilibrium state there will be no 
net flux of particles from the black hole, of course. 
However, for a black hole that formed from gravitational collapse 
\cite{manW97}, the thermal equilibrium state will settle down only 
asymptotically at large times, for the black hole will start to 
radiate only near and after the formation of the event horizon. If the 
universe is large enough, there will be a long time before infinity 
scatters the radiation back, and during this time there will be a net 
outgoing flux. With this in mind, we now want to calculate the black hole 
luminosity, i.e. that part of the total flux which is outgoing to infinity. 
To this end let us consider an outgoing Unruh--like mode, 
which near the horizon takes the form
\begin{equation}
\phi_{\lambda\omega} = (4\pi\omega)^{-1/2} r^{-1} e^{-i\omega U}
u_{\lambda}(x,y).
\label{out}
\end{equation}
The reason for considering this is that
to an external stationary observer, the 
collapse approach of the dust surface to the event horizon is 
exponentially fast in retarded time (this easily follows from the fact that 
the exterior metric is static all the time and equals the eternal 
black hole metric), i.~e.~we have for the radial coordinate $R$ of the
dust surface
\begin{eqnarray}
R(u)-r_+\simeq Ce^{-\kappa u}.
\end{eqnarray}
Thus the waves emitted from the surface of the dust appear 
enormously red--shifted with a continuously varying frequency of the 
form $\omega\exp(-\kappa u)$, which is just the phase of the Unruh mode 
(\ref{out}). Using (\ref{trans}) and (\ref{coeff}), we can express 
(\ref{out}) in the form
\begin{eqnarray}
\phi_{\lambda\omega} &=& \int_0^{\infty}\frac{d\omega'}
                  {\sqrt{4\pi\omega'}}r^{-1}
                  \bar{\gamma}_{\omega'\omega}e^{-i\omega'u}u_{\lambda}(x,y)
                  \nonumber \\ 
                  && - \int_0^{\infty}\frac{d\omega'}{\sqrt{4\pi\omega'}}r^{-1}
                  \epsilon_{\omega'\omega}e^{i\omega'u}u_{\lambda}(x,y),
\label{wavepack}             
\end{eqnarray}
Every component $\exp(-i\omega'u)$ in the wave packet (\ref{wavepack}) arrives 
at infinity as a mode
$\stackrel{\rightarrow}{T}_{\lambda}\!(\omega')\exp(-i\omega't+
i\Omega_{\lambda}'r_*)$, 
with $\Omega_{\lambda}' = \sqrt{\omega'^2 - \lambda^2/\ell^2}$. Note that only
waves with $\omega'> \lambda\ell^{-1}$ have oscillatory character when
they arrive at infinity, the others are damped exponentially.
(The fact that in adS space the ratio of angular momentum and energy
is limited above, is well--known, see e.~g.~\cite{avis78}). Having 
noted this, it is an easy matter to find the Bogoljubov coefficients 
relating the $|in\rangle$ to $|out\rangle$ vacuum. They are
\begin{eqnarray}
\alpha_{\lambda,\omega',\omega} &\equiv& \frac{\sqrt{\Omega'_{\lambda}}}
{\sqrt{\omega'}}\,\,\stackrel{\rightarrow}{\bar{T}}_{\lambda}\!(\omega')
\gamma_{\omega^{'},\omega} \nonumber \\ 
\beta_{\lambda,\omega',\omega} &\equiv& \frac{\sqrt{\Omega'_{\lambda}}}
{\sqrt{\omega'}}\,\,\stackrel{\rightarrow}{\bar{T}}_{\lambda}\!(\omega')
\epsilon_{\omega^{'},\omega}.
\end{eqnarray}
The relevant $\beta$--coefficients satisfy the relations
\begin{eqnarray}
\lefteqn{\int_0^{\infty}\bar{\beta}_{\lambda,\omega',\omega}
\beta_{\lambda,\omega^{''},\omega}\,d\omega=} \nonumber \\ 
&&\frac{\sqrt{\omega'^2 - \lambda^2\ell^{-2}}}{\omega^{'}}\,
\frac{|\!\stackrel{\rightarrow}{T}_{\lambda}\!(\omega')|^2}{e^{\omega^{'}/T}-1}
\,\,\delta(\omega^{'}-\omega^{''}),
\end{eqnarray}
where $T=\kappa/2\pi$. As $\delta(0)=T/2\pi$ for large time $T$, from this we 
conclude that the luminosity of the black hole is
\begin{equation}
L= \frac{1}{2\pi}\sum_{\lambda}g_{\lambda}\int_0^{\infty}
\omega^{-1}\sqrt{\omega^2 - \lambda^2\ell^{-2}}\,
\,|\!\stackrel{\rightarrow}{T}_{\lambda}(\omega)|^2
\frac{\omega\,d\omega}{e^{\omega/T}-1}. \label{lumi}
\end{equation}
As usual, the spectrum is not precisely planckian due to the presence 
of the grey body factor; however, it is only for large $\omega$ that it
approaches a form similar to that for the asymptotically flat
Sch\-warz\-schild black hole. We obtained the grey body factor
\begin{eqnarray}
\Gamma_{\lambda}(\omega)=\omega^{-1}\sqrt{\omega^2-\lambda^2\ell^{-2}}\,
|\!\stackrel{\rightarrow}{T}_{\lambda}\!(\omega)|^2,
\end{eqnarray} 
which means that a fraction
$|\!\stackrel{\rightarrow}{R}_{\lambda}\!(\omega)|^2=
1-\Gamma_{\lambda}(\omega)$ of the emitted 
particles can not reach infinity and is recaptured by the black hole 
(this fraction is also equal to 
$|\!\stackrel{\leftarrow}{R}_{\lambda}\!(\omega)|^2$,
the reflection coefficient 
for ingoing waves).\\
We may estimate $L$ in the
geometrical optics approximation, i.~e.~in the high frequency limit.
In this limit there is no reflection of emitted particles down 
the black hole if the inequality $\omega^2 > P_{\lambda}(r_{max})$ 
holds. In this case the transmission coefficient is proportional to
a step function, and it follows from Eq.~(\ref{unit}) that
\begin{equation}
|\!\stackrel{\rightarrow}{T}_{\lambda}\!(\omega)|^2=\omega(\omega^2-\lambda^2\ell^{-2})^{-1/2}
\theta(\omega^2-P_{\lambda}(r_{max})). \label{transm}
\end{equation}
From Eq.~(\ref{Pmax}), there is a $\lambda_{max}$ for which the inequality 
is true, which can be estimated to be
$\lambda_{max}(\omega)\simeq \omega\ell$ for 
large enough $\omega$. Also 
$\omega>\omega_0=3^{1/2}2^{-1}\ell^{-4/3}\eta^{1/3}$, for 
otherwise the inequality is violated for small $\lambda$. 
In the $g>1$ case, we also have $\lambda_{max}(\omega)\simeq \omega\ell$,
and $\omega > \omega_0$, with $\omega_0$ being identical to the value for
$g=1$ given above, provided $\eta > 0$ and $\eta \gg \ell$.
For $\eta > 0$ and $\eta \ll \ell$, or for $\eta \le 0$,
one has $\omega_0 = 0$. If the event horizon is spherical ($g=0$),
however, the situation changes. One now obtains for $\lambda_{max}(\omega)$
\begin{equation}
\lambda_{max}(\omega)^2 \simeq \frac{27\eta^2\ell^2\omega^2}
{\ell^2+27\eta^2}.
\end{equation}
In order to get the luminosity in the various cases, we
finally have to sum over $\lambda$ in (\ref{lumi}), at first sight a 
difficult task to perform since the degeneracy $g_{\lambda}$, where known, 
is a rather complicated expression. However, for compact manifolds 
there is the general Weyl's asymptotic formula \cite{chavel},
which in our case reads
\begin{equation}
\sum_{\lambda=0}^{\lambda_{max}}g_{\lambda} = \frac{{\cal A}_g}{4\pi}\,
\lambda_{max}^2 + {\cal O}(\lambda_{max}),
\end{equation}
where ${\cal A}_g$ is the area of a Riemann surface with genus $g$ and 
Gaussian curvature $K=-1$, $K=0$ or $K=1$ for $g>1$, $g=1$ or $g=0$,
respectively.  
So for a genus $g>1$ black hole, ${\cal A}_g=4\pi(g-1)$, for a 
torus ${\cal A}_1=\mathop{\rm Im}\nolimits\tau$, and for a sphere
${\cal A}_0=4\pi$.\\
Using this, we get for large $\lambda_{max}$ (which is 
fulfilled for $\omega$ sufficiently large, i.~e.~in the geometrical 
optics approximation) the estimate
\begin{equation}
L = \frac{{\cal A}_g}{8\pi^2}\int_{\omega_0}^{\infty}
\lambda_{max}(\omega)^2\frac{\omega\,d\omega}{e^{\omega/T}-1}.
\end{equation}
This yields for $g \ge 1$
\begin{equation}
L = C\pi^2{\cal A}_g\ell^2T^4, 
\label{StefBoltz}
\end{equation}
with
\begin{equation}
C \equiv \frac{4^{1/3}}{18}\int_1^{\infty}\frac{x^3\,dx}
{e^{4^{1/3}\pi x/\sqrt{3}}-1}\approx 0.0052,
\end{equation}
for $g=1$ or for $g>1$, $\eta>0$ and $\eta \gg \ell$.
For $g>1$ and  $0< \eta \ll \ell$, or for $g>1$ and $\eta \le 0$,
we get $C = 1/120$.\\
In the spherical case, the luminosity is
\begin{equation}
L = C\pi^2{\cal A}_0\frac{27\eta^2\ell^2}{\ell^2+27\eta^2}T^4,
\label{StefBoltzspher}
\end{equation}
with some numerical constant $C$, which we have not calculated explicitely
here.\\
Writing $L=-dM/dt$, and inserting the $M$--dependence $T \propto M^{1/3}$,
one derives an infinite lifetime for the toroidal black hole, in contrast
to the Sch\-warz\-schild case. (Of course, this is valid only in the
semiclassical limit. When the black hole mass approaches the Planck mass,
quantum gravity effects will occur). Therefore, if the universe is not 
too large to allow the black hole reaching the Planck mass before the 
radiation is reflected back, then, sooner or later, the black hole must 
start to grow until it reaches the temperature of the reflected 
radiation again. At this point the black hole should settle down to an 
equilibrium thermal state with a large entropy. In fact we have seen
that a wave like (\ref{out}) in a stationary anti-de Sitter black
hole, will propagate so as to become a positive frequency, ingoing 
wave on the future horizon relative to its canonical affine parameter.
This is the behaviour that marks the appearance of the thermal
equilibrium state and it means that an ingoing flux of energy enters
the black hole and balances the emitted, outgoing flux.\\ 
Inspecting (\ref{StefBoltz}) we observe that,
if we consider the black hole as a black body radiating with the Hawking
temperature, the area $A$ entering Stefan's law $L \propto AT^4$
is not the area of the event horizon, but an area determined by $\ell^2$,
i.~e.~by the cosmological constant! This is another intriguing feature
of topological black holes, different from asymptotically flat cases.
Note that the luminosity of a black hole with spherical event horizon
differs from that of the topological ones. Indeed, in the prefactor
of (\ref{StefBoltzspher}) also the parameter $\eta$ (which is equal to
the mass $M$ for $g=0$) enters, and in the limit $\ell \gg \eta$
(i.~e.~small cosmological constant) we recover
the known Schwarz\-schild result.\\
The found luminosity behaviour can be understood already
at a classical level, by
examining null geodesics in the black hole spacetime. We shall do this
in the following.
Using the fact that to every Killing vector there is an associated
constant of motion, for the radial coordinate $r$ one gets the equation
\begin{equation}
\frac{1}{2}\dot{r}^2 + \frac{1}{2}V(r)\frac{L^2}{r^2} = \frac{1}{2}E^2,
\label{classequ}
\end{equation}
where the dot denotes the derivative with respect to an affine
parameter. $V(r)$ is the square of the lapse function, $E$
is the constant of motion associated to $\partial_t$,
and $L^2=L_x^2+L_y^2$, where $L_x$ and $L_y$ belong to the Killing vectors
on the torus, namely
$\partial_x$ and $\partial_y$ respectively.
(For the Sch\-warz\-schild (AdS)
black hole, or for $g>1$, we limit ourselves to a fixed value of
$\theta$, e.~g.~$\theta=\pi/2$ for $g=0$, so
$L$ belongs to $\partial_{\phi}$. (Note that for a Riemann surface of genus
$g>1$, $\partial_{\phi}$ is a Killing vector only locally)).
Now the potential in
(\ref{classequ}) is given by
\begin{equation}
P(r) = V(r)\frac{L^2}{2r^2} = \left\{ \begin{array}{c@{\,,\quad}c}
          \frac{L^2}{2r^2}-\frac{ML^2}{r^3} & \mbox{Schwarzschild}\\ 
          \frac{L^2}{2r^2}(\delta_{g,1}-1) +
          \frac{L^2}{2\ell^2}-\frac{\eta L^2}{r^3} & g \ge 1
          \end{array} \right.
\end{equation}
As is well--known, in Sch\-warz\-schild spacetime, this potential
is zero at the horizon, has a maximum $P_{max}=L^2/54M^2$ at $r=3M$,
and then falls off to zero at infinity. Hence a particle coming
from infinity is captured by the black hole, if its "energy" $E^2/2$
exceeds the potential maximum. This means that the apparent impact
parameter $b \equiv L/E$ must be smaller then $\sqrt{27}M$ \cite{wald3},
and leads to the capture cross section
\begin{equation}
\sigma = \pi b^2 = 27\pi M^2
\end{equation}
for the Sch\-warz\-schild geometry. Thus a Sch\-warz\-schild black hole
absorbs like a black body with area $27\pi M^2$, a number directly
proportional to the horizon area. For the $g \ge 1$ black hole,
however, the situation is different. The potential $P(r)$ is also zero
at the horizon, but then increases monotonically to reach the constant
value $L^2/2\ell^2$ at infinity. Therefore every massless particle at
infinity with $E^2/2 > L^2/2\ell^2$, i.~e.~$b<\ell$, travelling towards
the black hole, is captured. This gives an absorption cross section
$\sigma = \pi\ell^2$, i.~e.~a $g \ge 1$ topological
black hole absorbs like a black
body with area $\pi\ell^2$, not like a black body with the event horizon
area. Of course, quantum mechanically there arises a local maximum
in the potential (see figure \ref{pot}), but this does not alter the
situation essentially.\\
For the $g=0$ Schwarz\-schild--AdS black hole,
$P(r)$ has a local maximum at $r=3M$, which leads to the capture
cross section
\begin{equation}
\sigma = \frac{27\pi M^2\ell^2}{\ell^2 + 27M^2},
\end{equation}
encountered already (if we identify $\eta=M$) in (\ref{StefBoltzspher}).
In this case, $\sigma$ is determined both by the cosmological constant
and the mass parameter, whereas for the $g \ge 1$ topological black
holes only the cosmological constant enters the capture cross section,
and thus the prefactor in the luminosity formula.

\section{Black Hole Spectrum and String States}

We have seen that an isolated black hole in anti--de Sitter space will 
ultimately settle down to a thermal equilibrium state with the Hawking 
temperature $T=\kappa/2\pi$ and some mass $M$. Such a black hole contributes 
to the total entropy its own entropy, $S_{bh}=A/4$. For a large mass 
black hole, the entropy and temperature depend on the mass as
\begin{eqnarray}
S=aM^{2/3},\qquad T=bM^{1/3},
\end{eqnarray}
with $a$ and $b$ computable constants (the two formulas are exact for 
the toroidal black hole, but only asymptotically correct for higher 
genus black 
holes and for $g=0$, which is the spherical anti--de Sitter black hole 
studied by Hawking--Page \cite{page83}). 
Hence the degeneracy of black hole states decreases in anti--de 
Sitter space, and the level density grows like
\begin{eqnarray}
\rho_{bh}(M)\simeq\exp(aM^{2/3}).
\end{eqnarray}
According to \cite{sanc93}, 
the same phenomenon occurs for strings in anti--de Sitter space  
where the level density at very large masses is 
($\ell^{-2}$ is proportional to the cosmological constant $\Lambda$)
\begin{eqnarray}
\rho_s(M)\simeq\exp(\sqrt{M\ell}).
\label{strden}
\end{eqnarray}
We now want to understand the black hole result by assuming a certain 
discrete spectrum for the black hole mass, with a certain degeneracy, 
and computing the corresponding partition function. The adiabatic 
invariant argument of Bekenstein would work in this case also, and 
suggests an area spectrum $A_n=\sigma n$, with $\sigma$ a number of order 
one \cite{mukh86,beke} (this result has been obtained also in loop quantum 
gravity \cite{rove95} and the membrane approach \cite{magg94}, for 
large quantum numbers). Then a mass spectrum arises of the form
\begin{eqnarray}
M_n=\alpha n^{3/2},
\end{eqnarray}
in sharp contrast with either the Schwarzschild or the string spectrum, 
$M_n\simeq\alpha\sqrt{n}$, or with the spectrum for strings 
in anti-de Sitter space, $M_n\simeq\ell^{-1}n$ for large 
$n$ \cite{sanc93}. The degeneracy will be assumed to be an increasing 
function $d(n)$, and the partition function takes the form
\begin{eqnarray}
Z=\sum_{n=0}^{\infty}e^{\ln d(n)}e^{-\beta\alpha n^{3/2}}.
\end{eqnarray}
We shall evaluate this quantity for large masses, i.e. small $\beta$, by 
using the steepest descent method. Replacing the sum with an integral 
we get
\begin{eqnarray}
Z=\int_0^{\infty}dt\,e^{\ln d(t)}e^{-\beta\alpha t^{3/2}}.
\end{eqnarray}
With $f(t)=\ln d(t)-\alpha\beta t^{3/2}$, the stationary point $t_0$, occurs at 
$f^{'}(t_0)=0$, and assuming also $f^{''}(t_0)>0$ we obtain the 
partition function
\begin{eqnarray}
\ln Z=\ln d_0-\left(\frac{d'_0}{d_0}\right)^3\frac{8}{27\alpha^2\beta^2}+
\frac{1}{2}\ln\left(\frac{2\pi}{f^{''}(t_0)}\right),
\end{eqnarray}
where $d_0=d(t_0)$. The partition function can also be computed in 
Euclidean quantum gravity for the asymptotically anti--de Sitter black 
holes \cite{page83,vanz97}, with the result
\begin{eqnarray}
\ln Z=\frac{4\pi^2\ell^4}{27\beta^2}.
\end{eqnarray}
This is exact for the toroidal black hole and valid approximatively 
for higher genus or spherical black holes. Comparing the two 
partition functions requires $d'/d=G$, 
$G$ being a constant, for a wide range of masses. In other words, 
$d(n)=G^n$ asymptotically for large $n$, with $G$ of order 
$\exp((\pi\alpha\ell^2)^{2/3})$ to match with the (robust) Euclidean 
result. \\ 
The obtained mass spectrum seems to be difficult to reconcile with string 
theory, even in anti--de Sitter space where $M\simeq\ell^{-1}n$,
asymptotically at mass level $n$ (in flat space this is 
$M\simeq\alpha^{'}\sqrt{n}$). 
On the other hand, the degeneracy of string states grows as $\exp\sqrt{n}$ 
(as in flat space), and therefore it is not obvious how the 
Susskind--Horowitz--Polchinski argument
\cite{suss93,horo97} should work. According to this argument, 
the black hole description breaks down when the horizon is of order 
the string scale, and the black hole becomes a highly excited string 
state. The mass of the black hole is $M_{bh}\simeq r_+^3G^{-1}\ell^{-2}$,
and the mass of a string state at level $n$ is $M_s\simeq\ell^{-1}n$, for 
$n\gg\ell/\ell_s$, where $\ell_s=\sqrt{\alpha'}$ is the string scale. 
The Newton constant is $G=g^2\alpha'$, where $g$ is the string 
coupling constant. Requiring the two masses to coincide (within a factor 
of order unity) when $r_+\simeq\ell_s$ gives  
\begin{eqnarray}
g^{-2}\simeq\ell\ell_s^{-1}n,
\end{eqnarray}
and the entropy is
\begin{eqnarray}
S=\frac{r_+^2}{4G}\simeq\frac{\ell_s^2}{4g^2\ell_s^2}=\ell\ell_s^{-1}n,
\end{eqnarray}
which disagrees with the string entropy $S\simeq\sqrt{n}$. This 
argument should be regarded as a very naive one. Strings in adS 
are not as well understood as in flat space and the mass formula is very 
complicated. For example, there is a regime where the mass--to--level
relation is as in flat space if $\ell/\ell_s\gg 1$. On the other 
hand, the cosmological constant is not a completely free parameter if  
string theory is to be anomaly free \cite{frad91}. In view of these 
facts, the correspondence principle of Horowitz and Polchinski can not be 
rejected on the above basis, but it remains to see how exactly it works.

\section{Summary and Discussion}

We have discussed quantum aspects of fields in the background of 
anti--de Sitter black holes. All the properties of them which are expected 
from the classical laws to the Euclidean approach are confirmed.\\ 
However, also new features emerge, which are related to the special 
asymptotic behaviour of anti--de Sitter space. Most surprising is the 
area dependence of the radiation formula (\ref{StefBoltz}), which is 
not determined by the area of the horizon. We also hope to have made clear 
that no Unruh--like states exist for eternal black holes.\\ 
In contrast, a 
black hole formed by collapse will radiate away its mass for a while 
after formation, until infinity will reflect it back. The black hole's 
temperature will then rise again up to the radiation temperature. At 
this point it should settle down to an equilibrium state at a certain 
Hawking temperature. Although we did not made efforts to compute the 
actual black hole evolution, this is a very reasonable picture, 
because anti--de Sitter space does not permit radiation to disperse to 
infinity. But then we have another version of the information loss 
paradox, because if the black hole does not completely evaporate there 
is no point for information to return.\\ 
The thermodynamical properties of anti--de Sitter black holes also lead 
to a peculiar mass spectrum, according to Bekenstein's view of the 
quantum structure of a black hole. This we have briefly discussed in 
relation to string theory, too. We think there is no 
simple way to understand the string--black hole correspondence 
principle in adS space, but we regard the question as unsettled for 
the time being, the point being that strings in adS behave 
very differently than in flat space.

\section*{Acknowledgement}

The part of this work due to D.K. has been
supported by a research grant within the
scope of the {\em Common Special Academic Program III} of the
Federal Republic of Germany and its Federal States, mediated 
by the DAAD.\\


\clearpage

\clearpage

\end{document}